\documentclass[sigconf]{acmart}
\AtBeginDocument{%
  \providecommand\BibTeX{{%
    \normalfont B\kern-0.5em{\scshape i\kern-0.25em b}\kern-0.8em\TeX}}}

\setcopyright{acmlicensed}
\copyrightyear{2018}
\acmYear{2018}
\acmDOI{XXXXXXX.XXXXXXX}

\acmConference[Conference acronym 'XX]{Make sure to enter the correct
  conference title from your rights confirmation emai}{June 03--05,
  2018}{Woodstock, NY}
\acmISBN{978-1-4503-XXXX-X/18/06}

\usepackage{algorithm}
\usepackage{algorithmic}
\usepackage{listings}
\usepackage{multirow}
\usepackage{float}
\newcommand{\lego}{\textsc{Legommenders}}
\newcommand{\legoicon}{\raisebox{-0.5ex}{\includegraphics[width=0.04\textwidth]{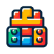}}}

\definecolor{softgreen}{RGB}{0, 128, 0} 
\definecolor{softred}{RGB}{230, 62, 62} 
\definecolor{softyellow}{RGB}{162, 162, 30} 

\newcommand{\coemedian}[1]{\textbf{\textcolor{softyellow}{#1}}}
\newcommand{\coelarge}[1]{\textbf{\textcolor{softred}{#1}}}

\definecolor{configkey}{RGB}{147, 29, 29}
\definecolor{configbool}{RGB}{24,103,30}
\definecolor{configint}{RGB}{24,70,103}
\newcommand{\cfgbool}[1]{\textcolor{configbool}{#1}}
\newcommand{\cfgint}[1]{\textcolor{configint}{#1}}
\newcommand{\cfg}[2]{\texttt{\textcolor{configkey}{#1}: #2}}

\begin{document}

\title[\lego{}]{\lego{} \legoicon{}: 
A Comprehensive Content-Based Recommendation Library with LLM Support}

\author{Qijiong Liu}
\affiliation{%
  \institution{The Hong Kong Polytechnic University}
  \country{Hong Kong SAR}
}
\email{liu@qijiong.work}

\author{Lu Fan}
\affiliation{%
  \institution{The Hong Kong Polytechnic University}
  \country{Hong Kong SAR}
}
\email{cslfan@comp.polyu.edu.hk}

\author{Xiao-Ming Wu}
\affiliation{%
  \institution{The Hong Kong Polytechnic University}
  \country{Hong Kong SAR}
}
\email{xiao-ming.wu@polyu.edu.hk}

\renewcommand{\shortauthors}{Liu and Wu, et al.}

\begin{abstract}

We present \lego{}, a unique library designed for content-based recommendation that enables the joint training of content encoders alongside behavior and interaction modules, thereby facilitating the seamless integration of content understanding directly into the recommendation pipeline. \lego{} allows researchers to effortlessly create and analyze over 1,000 distinct models across 15 diverse datasets. Further, it supports the incorporation of contemporary  large language models, both as feature encoder and data generator, offering a robust platform for developing state-of-the-art recommendation models and enabling more personalized and effective content delivery.

\end{abstract}





\maketitle

\section{Introduction}


In online content discovery, recommender systems play a pivotal role as navigators, significantly enhancing user experiences through personalized content delivery. Traditionally, recommender systems have predominantly relied on transductive learning mechanisms~\cite{ctr-dcn,ctr-deepfm}. This approach utilizes static user and item identifiers (IDs) to generate predictions based on existing data. While effective within the confines of known datasets, this method presents limitations. It struggles to adapt to new users and items, often referred to as the ``cold start'' problem, and is less responsive to shifts in user preferences over time.


Modern recommender systems have shifted from transductive learning to inductive learning, utilizing inherent content features and user historical behaviors to create more dynamic models~\cite{model-naml,model-plmnr,model-uist,model-cost,model-store,model-iisan,model-embsum,model-spar,model-greenrec,model-mm4rec}. Specifically, a content-based recommendation model typically consists of three components: (1) A \emph{content operator} that generates embeddings for both the candidate item and each item in the user's behavior sequence. (2) A \emph{behavior operator} that fuses the user sequence into a unified user embedding. (3) A \emph{click predictor} that calculates the click probability for the user on the given item. Notably, the content operator can either be trained jointly with the other two modules or be decoupled from them.


However, most existing recommender system libraries~\cite{bm-recbole-2.0,bm-fuxictr,bm-ducho-2} employ a decoupled design, which fails to adapt the content encoder to specific recommendation scenarios. Typically, item embeddings are generated by a \emph{pretrained} content encoder and used as an initial step. Although this approach enhances model efficiency, it has a significant limitation: the pretrained embeddings are often too general and not well-aligned with the specific recommendation context, leading to suboptimal recommendations.

\begin{figure}
    \centering
    \includegraphics[width=.8\linewidth]{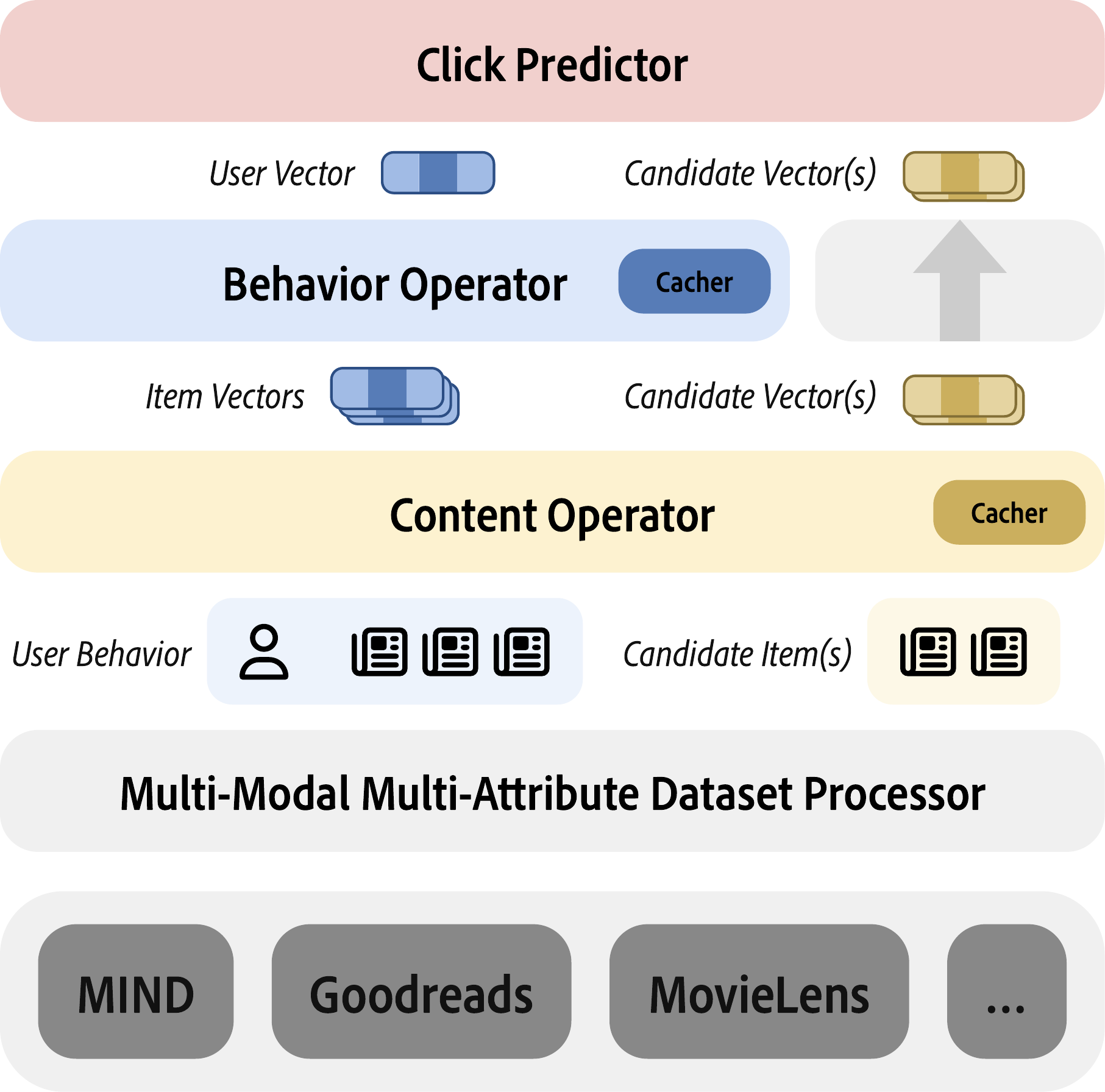}
    \caption{Overview of the \lego{} package.}
    \label{fig:overview}
\end{figure}

\begin{table*}[t]
\caption{Comparison to exisiting recommendation benchmarks (\textcolor{softgreen}{\checkmark} | \coemedian{--} | \coelarge{$\times$} means totally | partially | not met, respectively). ``Partially met'' indicates
incomplete availability.}\label{tab:benchmark-comparison}

\begin{tabular}{lcccccccc}
\toprule
\textbf{Feature} & \textbf{TorchRec} & \textbf{DeepCTR} & \textbf{DeepRec} & \textbf{RecBole} & \textbf{FuxiCTR} & \textbf{BARS} & \textbf{Ducho} & \textbf{\lego{}} \\
Year & (\citeyear{bm-torchrec}) & (\citeyear{bm-deepctr}) & (\citeyear{bm-deeprec}) & (\citeyear{bm-recbole-2.0}) & (\citeyear{bm-fuxictr}) & (\citeyear{bm-bars}) & (\citeyear{bm-ducho-2}) & \textit{(ours)} \\
\midrule
Content-based     & \coelarge{$\times$} & \coelarge{$\times$} & \coelarge{$\times$} & \coemedian{--}     & \coemedian{--}     & \coemedian{--}  & \textcolor{softgreen}{\checkmark}  & \textcolor{softgreen}{\checkmark} \\
LLM Encoding & \coelarge{$\times$} & \coelarge{$\times$} & \coelarge{$\times$} & \coelarge{$\times$} & \coelarge{$\times$} & \coelarge{$\times$} & \textcolor{softgreen}{\checkmark}  & \textcolor{softgreen}{\checkmark} \\
End-to-end Training   & \coelarge{$\times$} & \coelarge{$\times$} & \coelarge{$\times$} & \coelarge{$\times$} & \coelarge{$\times$} & \coelarge{$\times$} & \coelarge{$\times$} & \textcolor{softgreen}{\checkmark} \\
Fast Evaluation       & \coelarge{$\times$} & \coelarge{$\times$} & \coelarge{$\times$} & \coelarge{$\times$} & \coelarge{$\times$} & \coelarge{$\times$} & \coelarge{$\times$} & \textcolor{softgreen}{\checkmark} \\
\# Models   & <10 & 30+ & 10+ & 150+ & 50+ & 50+ & <10 & 1000+ \\
\bottomrule
\end{tabular}
\end{table*}


In contrast, our \lego{} library offers a unique and innovative feature by enabling the joint training of content operators alongside other modules. This capability allows for the seamless integration of content understanding directly into the recommendation pipeline. As shown in Figure~\ref{fig:overview}, \lego{} comprises four core components: the dataset processor, content operator, behavior operator, and click predictor. By combining these built-in operators and predictors in the configuration files, researchers can design over \textbf{1,000} recommendation models, utilizing 15 content operators, 8 behavior operators, and 9 click predictors. Remarkably, \textbf{95\%} of these models have not been tested or published before. These models can be evaluated across multiple content-based recommendation scenarios using over \textbf{15} datasets. Moreover, \lego{} is the \emph{first} library to inherently support large language models (LLMs). It not only accepts augmented data generated by LLMs for training but also integrates open-source LLMs as content operators/encoders. This dual capability allows \lego{} to improve data quality and generate superior data embeddings using LLMs.


\lego{} is designed to offer researchers and practitioners a comprehensive, flexible, and user-friendly platform for conducting experiments and analyses in content-based recommendation in the era of LLMs, aiming to facilitate new research directions in the field. The \lego{} library, including code, data, and documentation, is accessible at: \url{https://github.com/Jyonn/Legommenders}.
\section{Comparison to Existing Benchmarks}



Despite the success of previous research, there remains a significant lack of standardized benchmarks and uniform evaluation protocols for content-based recommendation systems.
As summarized in Table~\ref{tab:benchmark-comparison}, traditional recommendation libraries typically accept \emph{only ID-based} features and do not utilize large language models (LLMs) for content encoding. The recent library Ducho~\cite{bm-ducho-2} offers multimodal feature extraction for downstream recommendation models but maintains a decoupled design. In contrast, \lego{} is currently the only library that supports end-to-end training of content operators, behavior operators, and click predictors. Furthermore, we have developed an inference caching pipeline that achieves up to a 50x speedup in evaluation. By enabling easy modular combinations, we provide over 1,000 models, which is six times more than the largest existing model libraries.

\section{\lego{}: Details and Usage}

\begin{figure}[!h]
    \centering
    \includegraphics[width=.9\linewidth]{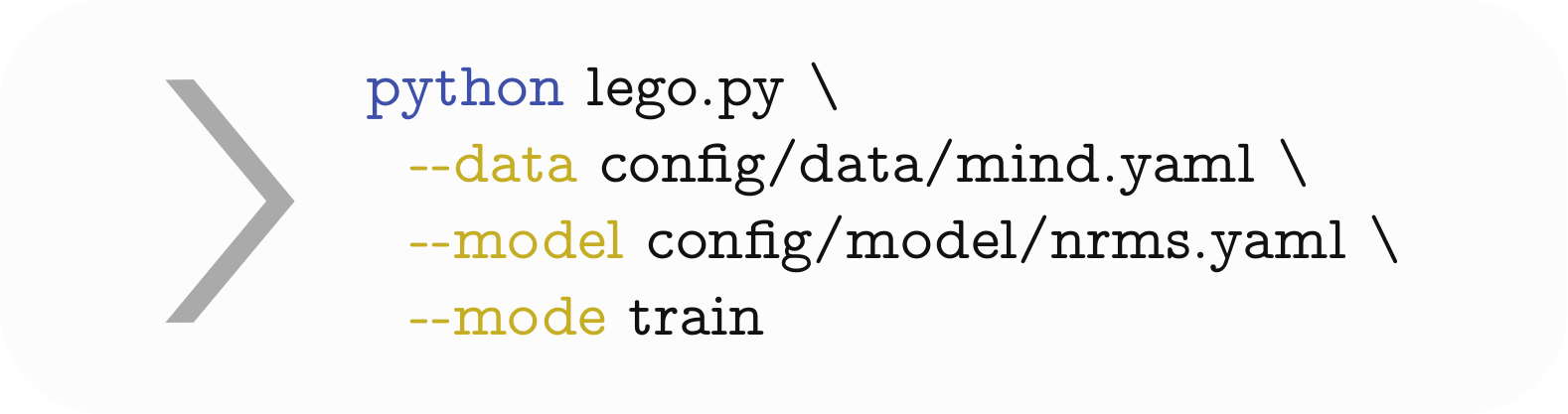}
    \caption{A quick use of \lego{}.}
    \label{fig:use}
\end{figure}

In this section, we first introduce the supported recommendation tasks, followed by a detailed discussion of the key components. We then present our caching pipeline and conclude with an overview of the algorithm flow.

\subsection{Recommendation Tasks}

The \lego{} library supports two fundamental recommendation tasks: \textbf{matching} and \textbf{ranking}.

In the matching task, given a user and a set of \( K+1 \) candidate items (one positive and \( K \) negative), the model performs a \( K+1 \) classification task to identify the positive item, which is formulated as:
\[
\hat{y}_{ui} = \text{softmax}(f(x_u, x_i)),
\]
where \( f(x_u, x_i) \) is mostly a simple dot operation, calculates the user-item feature relevance. The training objective is to maximize the likelihood of the correct positive item:
\[
\mathcal{L}_{\text{matching}} = - \sum_{(u, i) \in \mathcal{D}} \sum_{i=1}^{K+1} y_{ui} \log(\hat{y}_{ui}),
\]
where \( y_{ui} \) is the binary label for item \( i \), and \( \mathcal{D} \) is the dataset of user-item pairs.

\begin{figure}
    \centering
    \includegraphics[width=\linewidth]{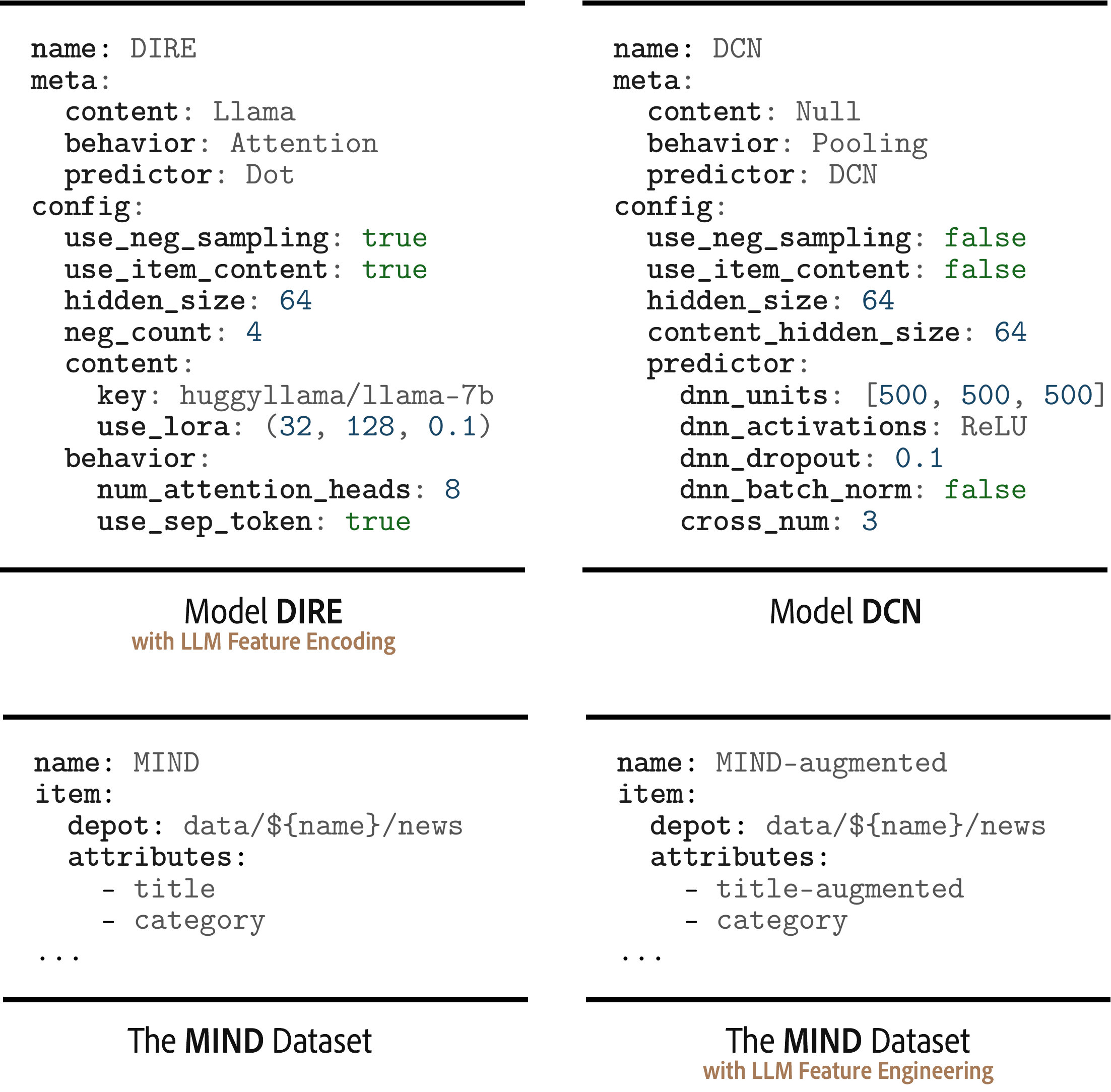}
    \caption{Examples for model and dataset configurations.}
    \label{fig:configurations}
\end{figure}

In the ranking task, the model predicts the click probability for a given user-item pair, denoted as: 
\[
\hat{r}_{ui} = f(x_u, x_i),
\]
where \( f(x_u, x_i) \) can be deep CTR models and trained to minimize the mean squared error between the predicted and actual labels:
\[
\mathcal{L}_{\text{ranking}} = \frac{1}{|\mathcal{D}|} \sum_{(u, i) \in \mathcal{D}} (y_{ui} - \hat{r}_{ui})^2.
\]
The task can be configured with the \cfg{use\_neg\_sampling}{\cfgbool{false}} and \cfg{neg\_count}{\cfgint{0}} parameters in the model configuration file, as depicted in Figure~\ref{fig:configurations}.

\subsection{Dataset Processor}

\lego{} provides a range of built-in processors for various recommendation scenarios, including news recommendation (MIND~\cite{ds-mind}, PENS~\cite{ds-pens}, Adressa~\cite{ds-adressa}, and Eb-NerD~\cite{ds-ebnerd}), book recommendation (Goodreads~\cite{ds-goodreads} and Amazon Books~\cite{ds-amazon}), movie recommendation (Movielens~\cite{ds-movielens} and Netflix~\cite{ds-netflix}), and music recommendation (Amazon CDs~\cite{ds-amazon} and Last.fm~\cite{ds-lastfm})
Dataset-specific processors convert the data into a unified format using the UniTok library\footnote{\url{https://pypi.org/project/UniTok/}}, which includes an item content table, a user behavior table, and an interaction table. 

In contrast to FuxiCTR~\cite{bm-fuxictr} and BARS~\cite{bm-bars}, which merge multiple tables into one, \lego{} adheres to the second normal form, significantly reducing data redundancy. This decoupled storage design allows \lego{} to easily accommodate augmented data generated by large language models simply by modifying the selected item attributes, as shown in Figure~\ref{fig:configurations}.

\subsection{Content Operator}

\lego{} decomposes content-based recommendation models into the content operator, behavior operator, and click predictor. This modular design allows for flexible selection and combination of these components to create new recommenders. The built-in content operators include average pooling used by text-based CTR models~\cite{ctr-dcn,ctr-deepfm}, convolutional neural networks (CNN)~\cite{model-cnn} used by NAML~\cite{model-naml} and LSTUR~\cite{model-lstur}, Attention~\cite{model-attention} used by NRMS~\cite{model-nrms} and PREC~\cite{model-prec}, and Fastformer~\cite{model-fastformer}.

\lego{} supports the use of LLMs as content operators, such as BERT~\cite{model-bert}, LLaMA~\cite{model-llama1}, and other open-source models available on Hugging Face\footnote{\url{https://huggingface.co}}. Building on insights from previous works~\cite{model-plmnr,model-prec,model-once}, we propose a training method that freezes the lower layers while fine-tuning the upper layers, including LoRA-based PEFT~\cite{lora}. This approach achieves up to 100x training acceleration compared to full fine-tuning, as it requires only the parameter ``\texttt{----mode split ----layer <N>}''.

Additionally, \lego{} is compatible with identifier-based recommenders by setting \cfg{use\_item\_content}{\cfgbool{false}} and using a randomly initialized item embedding table, such as DCN~\cite{ctr-dcn} model in Figure~\ref{fig:configurations}. It is compatible with decoupled content operator designs, like Ducho~\cite{bm-ducho-2}, by setting \cfg{use\_item\_content}{\cfgbool{false}} and using an embedding table from pretrained models.

\subsection{Behavior Operator and Click Predictor}



The built-in behavior operators encompass several mechanisms, including average pooling, which is utilized by CTR models; Additive Attention~\cite{model-ada}, employed by NAML~\cite{model-naml}; GRU~\cite{model-gru}, used by LSTUR~\cite{model-lstur}; and Attention, which is implemented in NRMS~\cite{model-nrms}, BST~\cite{model-bst}, and PLM-NR~\cite{model-plmnr}. Additionally, PolyAttention is utilized by MINER~\cite{model-miner}, among others. The built-in click predictors include the dot product, a method widely used in numerous matching-based models and various CTR models~\cite{ctr-dcn,ctr-gdcn,ctr-deepfm}, which rely on feature interaction modules as their core design.

\subsection{Caching Pipeline}

During inference and evaluation, the model parameters are fixed. Traditional recommendation libraries and content-based recommender systems dynamically encode user and item embeddings for each user-item pair in the test set and calculate the click probability in real-time. This results in redundant computations, which can be exacerbated by the cascaded design of content and behavior operators. To mitigate this, we propose content and behavior cachers, as illustrated in Figure~\ref{fig:overview}, which precompute and store embeddings for all items and users during the inference phase. During subsequent inferences, only the lightweight click predictor is required. The acceleration gained from caching becomes more pronounced as the frequency of repeated users and items, as well as with the sizes of the content and behavior operators. In some cases, this approach can yield up to 50x inference speedup. The caching mechanism is enabled by default and seamlessly integrated into the recommendation model, demonstrated in Algorithm~\ref{algo:lego}.

\begin{algorithm}[t]
\caption{Python-style Code for Training and Inference}
\label{algo:lego}
\begin{lstlisting}[frame=none]
class Legommenders:
    def forward(self, user, item, labels):
        content_op = self.content_cacher
        if self.content_op and self.training:
            content_op = self.content_op
        user = content_op(user)
        item = content_op(item)

        behavior_op = self.behavior_cacher
        if self.training:
            behavior_op = self.behavior_op
        user = behavior_op(user)

        scores = self.predictor(user, item)
        if self.training:
            return self.loss_fct(scores, labels)
        return scores
\end{lstlisting}
\end{algorithm}
\section{Experiments}


In this section, we present a selection of benchmark results for representative models on the MIND dataset. These results illustrate the robust modular composition capabilities of \lego{} and its support for LLMs. 

\begin{table}[]

\centering
\caption{Selected benchmark results on the MIND dataset. ``ContentOp'' and ``BehaviorOp'' denote Content Operator and Behavior Operator, respectively. ``Original'' and ``Augmented'' refer to the original MIND dataset and the GPT-augmented dataset as in ONCE~\cite{model-once}. Null(Llama1) indicates the decoupled design using Llama for item embedding extraction.}
\label{tab:exp}

\setlength\tabcolsep{1pt}

\resizebox{\linewidth}{!}{
\begin{tabular}{c|cccc|ccc}
\toprule
\textbf{Dataset} & \textbf{Model} & \textbf{ContentOp} & \textbf{BehaviorOp} & \textbf{Predictor} & \textbf{AUC} & \textbf{MRR} & \textbf{N@5} \\
\midrule
Original & NAML$_\text{ID}$ & Null & AdditiveAttention & Dot & 50.13 & 23.01 & 22.35 \\
Original & DCN & Null & Pooling & DCN & 53.92 & 25.18 & 24.43 \\
Original & DIN & Null & Null & DIN & 55.95 & 25.88 & 25.95 \\
\midrule
Original & DCN$_\text{text}$ & Pooling & Pooling & DCN & 62.63 & 29.73 & 30.52 \\
Original & DIN$_\text{text}$ & Pooling & Null & DIN & 62.90 & 30.06 & 30.65 \\
Original & NAML & CNN & AdditiveAttention & Dot & 61.75 & 30.60 & 31.35 \\
Original & NRMS & Attention & Attention & Dot & 61.71 & 30.20 & 30.98 \\
Original & MINER & BERT & PolyAttention & Attention & 63.88 & 32.19 & 33.04 \\
Original & Fastformer & Fastformer & Fastformer & Dot & 62.26 & 31.14 & 31.90 \\
\midrule
Original & PLM-NR & BERT & Attention & Dot & 64.08 & 31.24 & 32.35 \\
Original & DIRE & Llama1 & Attention & Dot & 68.50 & 36.21 & 38.11 \\
Original & DIRE & Null(Llama1) & Attention & Dot & 68.10 & 35.33 & 36.91 \\
\midrule
Augmented & DCN$_\text{text}$ & Pooling & Pooling & DCN & 65.77 & 32.86 & 34.10 \\
Augmented & NAML & CNN & AdditiveAttention & Dot & 63.88 & 32.17 & 33.14 \\
Augmented & NRMS & Attention & Attention & Dot & 63.71 & 32.14 & 33.11 \\
Augmented & PLM-NR & BERT & Attention & Dot & 65.13 & 32.98 & 34.30 \\
Augmented & ONCE & Llama1 & Attention & Dot & 68.74 & 36.66 & 38.60 \\
\bottomrule
\end{tabular}
}
\end{table}

All baselines use the same hyperparameters, including embedding dimension, number of attention heads, learning rate, and others, to ensure consistency. Due to space limitations, we will provide the full configurations in our repository for reproducibility. The results show that: 1) models trained on GPT-augmented datasets consistently outperform those using the original datasets; 2) baselines incorporating more complex language models tend to achieve better performance; 3) LLM-finetuning scheme outperforms the decoupled design. These findings highlight the strong content understanding capabilities of LLMs, further underscoring the contribution of our library.

\section{Conclusion}

We have introduced \lego{}, a library designed for content-based recommendation systems, which stands out due to its ability to jointly train content operators, behavior operators, and click predictors for inductive learning, its modular design, and its support for LLMs as both content encoders and data generators. We believe that \lego{} will serve as a valuable tool, significantly accelerating research within the recommendation community.


\bibliographystyle{ACM-Reference-Format}
\bibliography{Legommenders}

\end{document}